\documentclass[twocolumn,amsmath,amssymb,prl]{revtex4}

\usepackage{graphicx}
\usepackage{dcolumn}
\usepackage{bm}
\usepackage{color}
\usepackage{notes2bib}

\RequirePackage[normalem]{ulem} 
\RequirePackage{color}\definecolor{RED}{rgb}{1,0,0}\definecolor{BLUE}{rgb}{0,0,1} 
\providecommand{\DIFaddbegin}{} 
\providecommand{\DIFdelbeginFL}{} 

\begin{document}

\bibliographystyle{apsrev}

\title{Excitation and coherent control of spin qudit modes with sub-MHz  spectral resolution}

\author{V.~A.~Soltamov$^{1,2}$}
\author{C.~Kasper$^{2}$}
\author{A.~V.~Poshakinskiy$^{1}$}
\author{A.~N.~Anisimov$^{1}$}
\author{E.~N.~Mokhov$^{1,3}$}
\author{A.~Sperlich$^{2}$}
\author{S.~A.~Tarasenko$^{1}$}
\author{P.~G.~Baranov$^{1,4}$}
\author{G.~V.~Astakhov$^{2,1}$}
\email[E-mail:~]{astakhov@physik.uni-wuerzburg.de} 
\author{V.~Dyakonov$^{2}$}
\email[E-mail:~]{dyakonov@physik.uni-wuerzburg.de}

\affiliation{$^1$Ioffe Institute, 194021 St.~Petersburg, Russia  \\ 
$^2$Experimental Physics VI, Julius-Maximilian University of W\"{u}rzburg, 97074 W\"{u}rzburg, Germany \\ 
$^3$National Research University of Information Technologies, Mechanics and Optics, St Petersburg, Russia., 197101 St. Petersburg, Russia  \\
$^4$Peter the Great St.~Petersburg Polytechnic University, St.~Petersburg, 195251, Russia }

\begin{abstract}  
Quantum bit or qubit is a two-level system, which builds the foundation for quantum computation, simulation, communication and sensing \cite{Awschalom:2013in}. Quantum states of higher dimension, i.e., qutrits (D~=~3) and especially qudits (D~=~4 or higher),  offer significant advantages.  
Particularly, they can provide noise-resistant quantum cryptography \cite{Kaszlikowski:2003js}, simplify  quantum logic \cite{Lanyon:2008gvc} and improve quantum metrology \cite{Shlyakhov:2018dg}. Flying and solid-state qudits have been implemented on the basis of photonic chips \cite{Kues:2017db} and superconducting circuits  \cite{Neeley:2009ic}, respectively.  However, there is still a lack of room-temperature qudits with long coherence time and high spectral resolution. The silicon vacancy centers in silicon carbide (SiC) with spin \textit{S}~=~3/2 are quite promising in this respect \cite{Riedel:2012jq}, but until now they were treated as a canonical qubit system \cite{Mizuochi:2002kl,Kraus:2013di,Widmann:2014ve,Carter:2015vc,Simin:2017iw}. Here, we apply a two-frequency protocol to excite and image multiple qudit modes in a SiC spin ensemble under ambient conditions. Strikingly, their spectral width is about one order of magnitude narrower than the inhomogeneous broadening of the corresponding spin resonance. By applying Ramsey interferometry to these spin qudits, we achieve a spectral selectivity of 600~kHz and a spectral resolution of 30~kHz.  As a practical consequence, we demonstrate absolute DC magnetometry insensitive to thermal noise and strain fluctuations.
\end{abstract}

\date{\today}

\maketitle

In our experiments, we focus on the silicon vacancies ($\mathrm{V_{Si}}$) in two main hexagonal polytypes of SiC, namely 6H and 4H, with natural and modified isotope abundance. The 6H- and 4H-$^{28}$SiC samples, purified from the $^{29}$Si isotope, were grown by the seeded physical vapour transport method \cite{Tairov:1978hd}. The $^{29}$Si isotope abundance of 1\%, as established by the electron paramagnetic resonance (EPR) spectroscopy (Supplementary Fig.~S1), is 4.5 times smaller than that in natural SiC crystals.

The $\mathrm{V_{Si}}$ centers in SiC possess a specific spin $S = 3/2$ fine structure \cite{Simin:2016cp}.  Compared to comon spin-$1/2$ qubits, these centers provide additional functionality for quantum applications because they can be in multiple superpositions of four basis states in the Hilbert space with the spin projections $m_S = \pm 3/2, \pm 1/2$. Similar to atomic orbitals, an ensemble of such spin qudits can be described by 15 linearly independent spherical multipoles, with three components being the spin dipole $\mathcal{P}$, five components being the spin quadrupole  $\mathcal{D}$ and seven components being the spin octupole  $\mathcal{F}$ \cite{Tarasenko:2017ky}. In the simplest case, when the external magnetic field $\mathbf{ B}$ is applied along the symmetry axis  ($c$-axis of the SiC lattice), the density matrix $\rho$ is diagonal and can be written in the form $ \rho = \mathcal{I} / 4 + p_0 \mathcal{P}_0 + d_0 \mathcal{D}_0+ f_0 \mathcal{F}_0$\DIFaddbegin. Here, $\mathcal{I}$, $\mathcal{P}_0$, $\mathcal{D}_0$, and $\mathcal{F}_0$ are linearly-independent $4 \times 4$ diagonal matrices (table~\ref{Multipole}). To describe our experimental results, we use all (15 + unit $\mathcal{I}$) basis  matrices (Supplementary Theory). When qudit multipoles are excited, they decay independently, and in the spherical approximation, there are only three relaxation times: that of the spin dipole ($T_p$), quadrupole ($T_d$), and octupole ($T_f$) (table~\ref{Multipole}). 

\begin{table}[bth]
\caption{Illustration of the $\mathrm{V_{Si}}$ spin multipoles and their mathematic representation via the basis diagonal matrices  $\mathcal{I}$, $\mathcal{P}_0$, $\mathcal{D}_0$  and $\mathcal{F}_0$\DIFdelbeginFL . }
\begin{center}
\begin{tabular}{|c|c|c|c|c|}
Qudit & Thermal & Spin  & Spin & Spin   \\
modes: & equilibrium & dipole  & quadrupole & octupole   \\
Density  & \includegraphics[width=.09\textwidth]{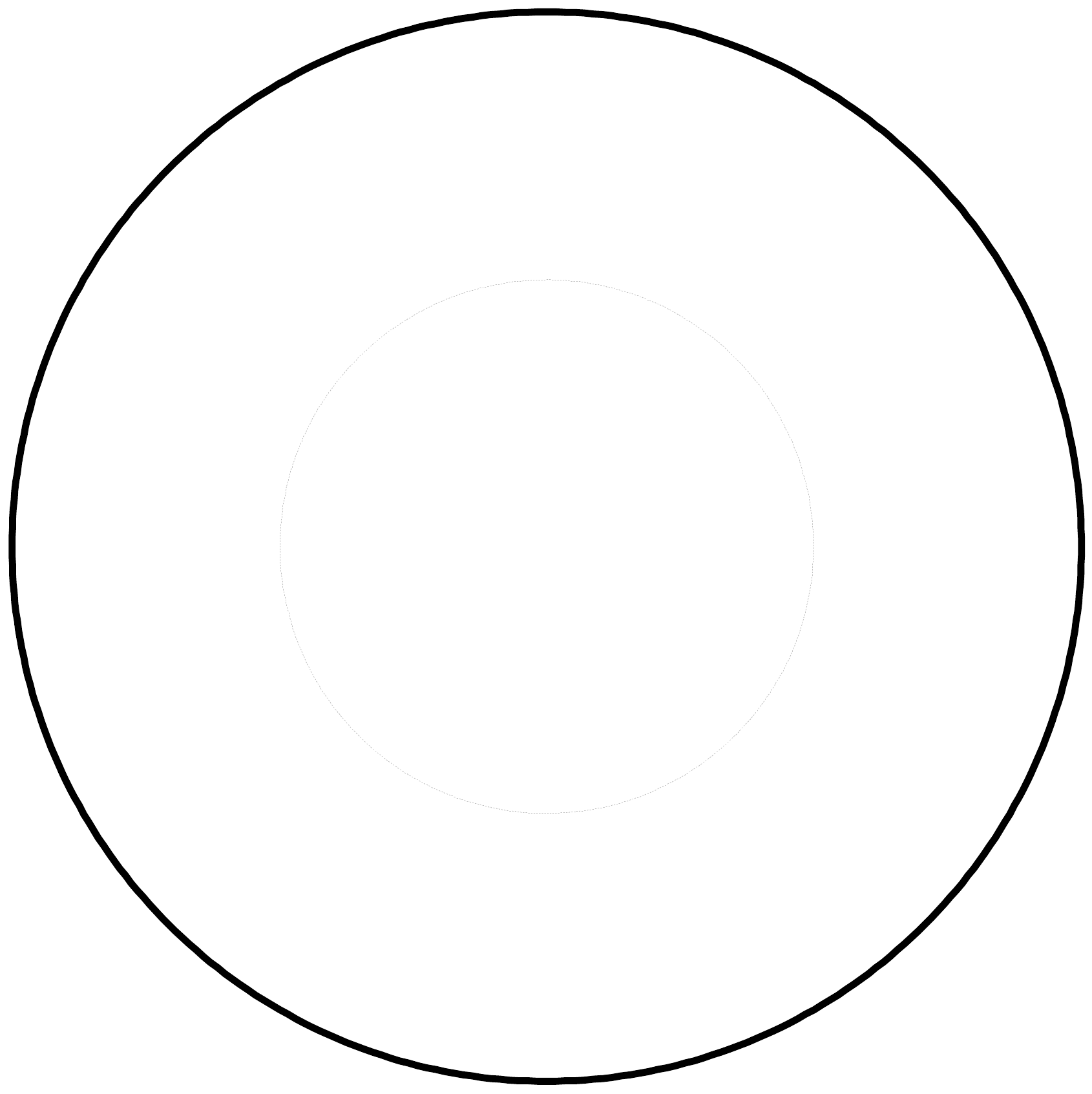} & \includegraphics[width=.09\textwidth]{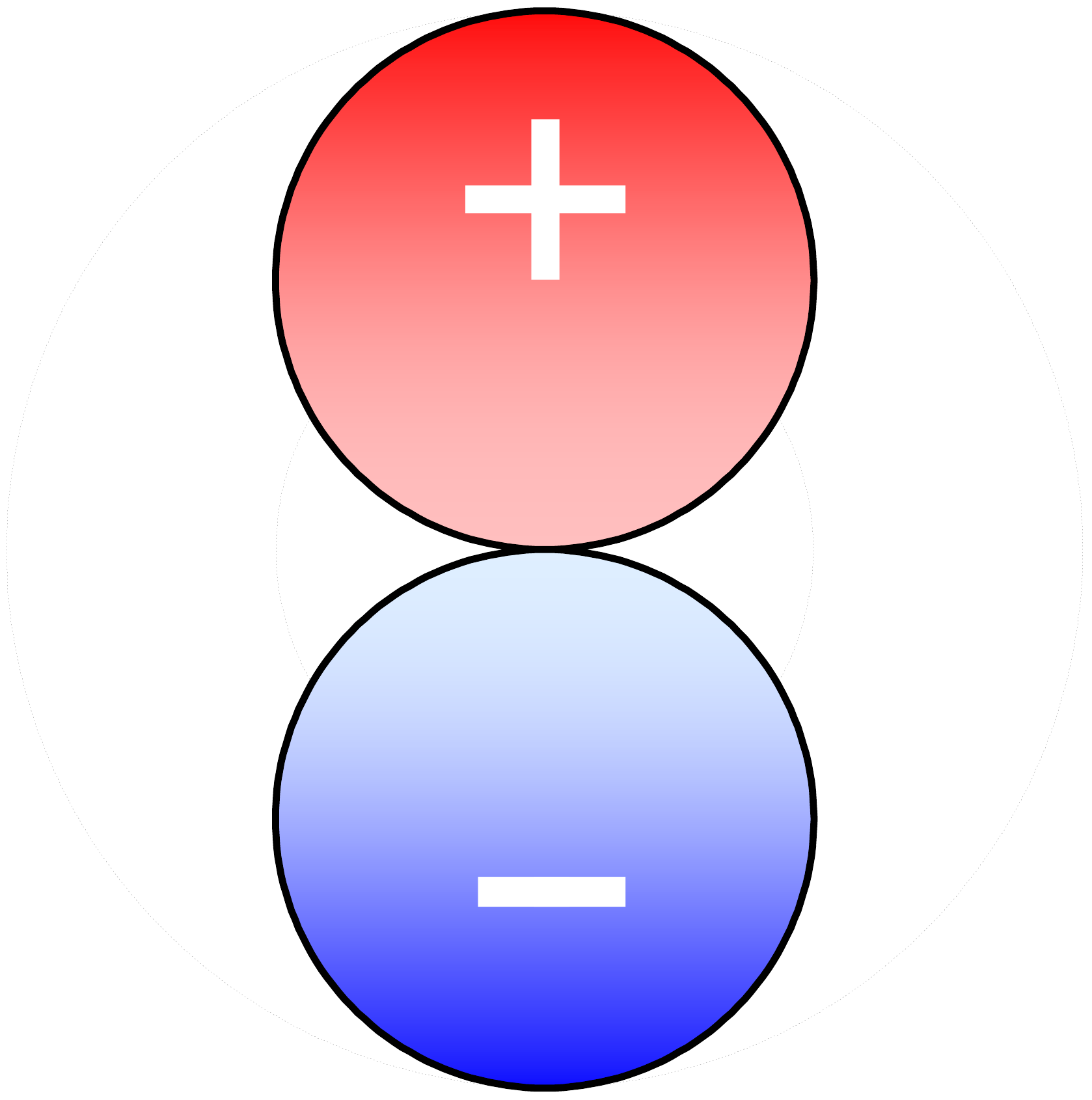} & \includegraphics[width=.09\textwidth]{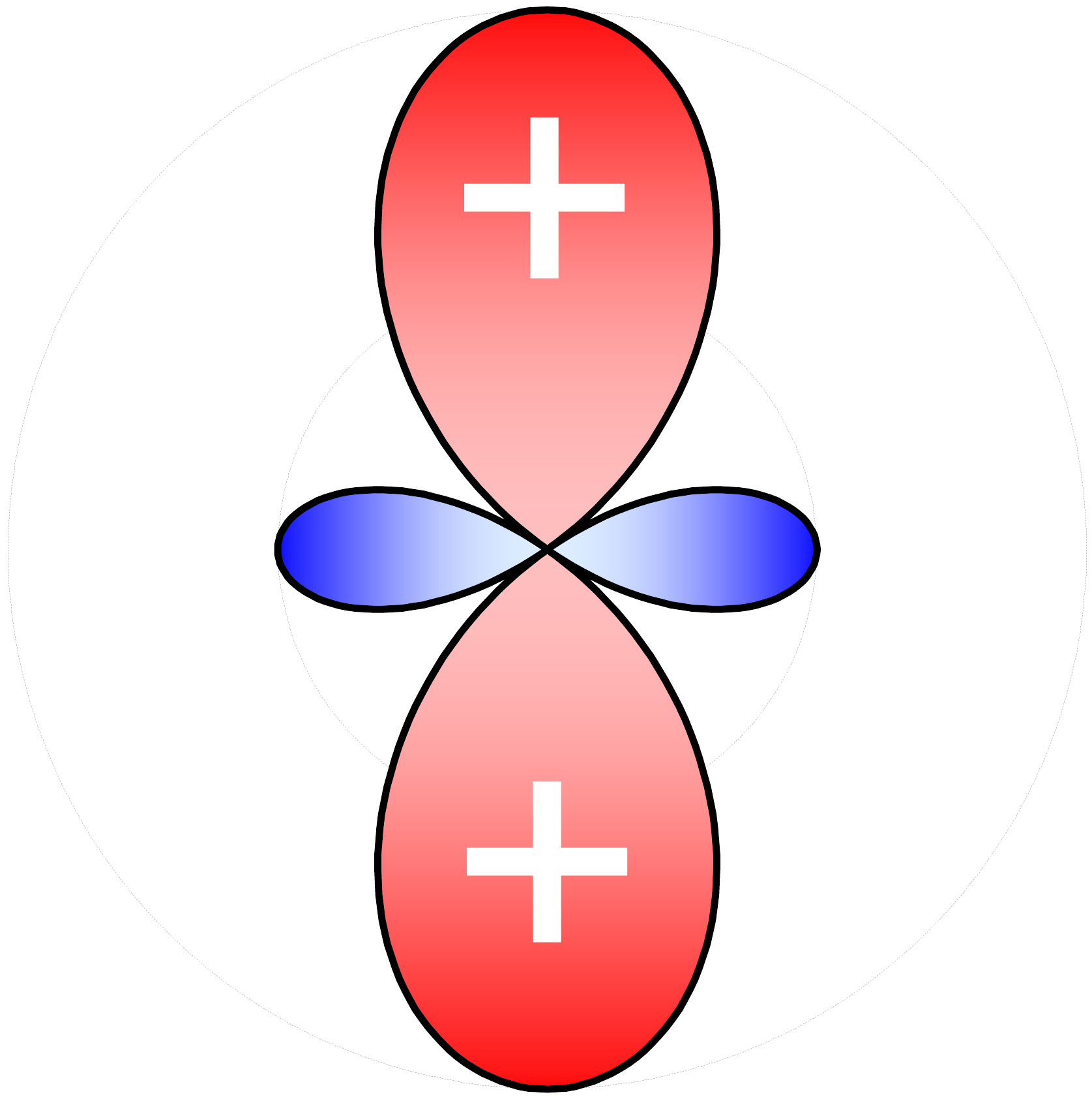} & \includegraphics[width=.09\textwidth]{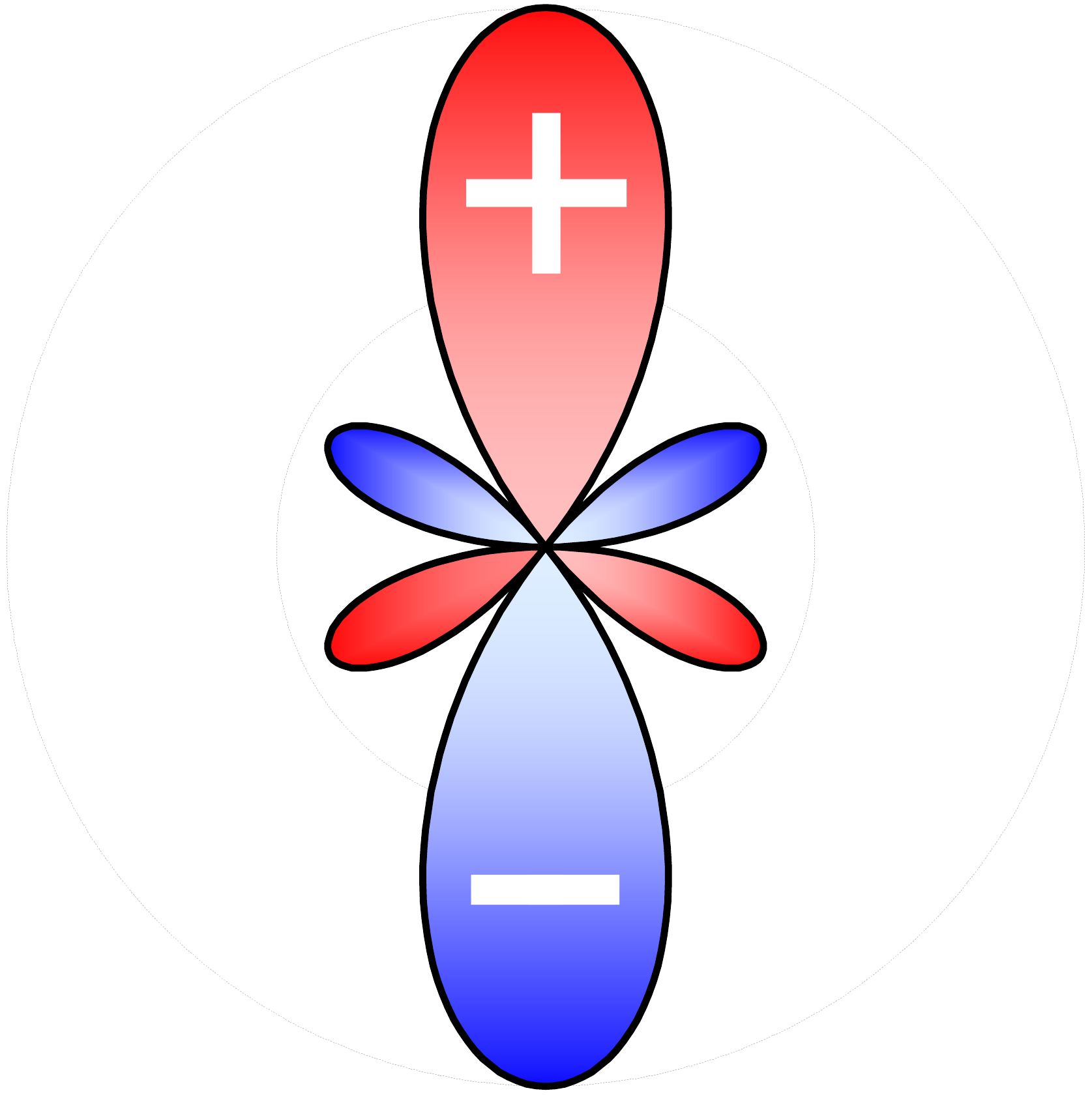}   \\
$\downarrow$ matrix  &  &  & &    \\
\hline
$\rho_{+\frac{3}{2},+\frac{3}{2}} \hspace{-1mm} =$ & $1/4$ & $+ 3 p_0 / \sqrt{20}$ & $+ d_0 / 2$ & $+ f_0 / \sqrt{20}$   \\
$\rho_{+\frac{1}{2},+\frac{1}{2}} \hspace{-1mm}=$ & $1/4$ & $+ p_0 / \sqrt{20}$ & $- d_0 / 2$ & $- 3 f_0 / \sqrt{20}$   \\
$\rho_{-\frac{1}{2},-\frac{1}{2}} \hspace{-1mm}=$ & $1/4$ & $- p_0 / \sqrt{20}$ & $- d_0 / 2$ & $+ 3 f_0 / \sqrt{20}$   \\
$\rho_{-\frac{3}{2},-\frac{3}{2}} \hspace{-1mm} =$ & $1/4$ & $- 3 p_0 / \sqrt{20}$ & $+ d_0 / 2$ & $- f_0 / \sqrt{20}$   \\
\hline
 $\,\,\,\,\,\,\,  \rho (t) = $ & $ \mathcal{I} / 4$ & $p_0  \mathcal{P}_0 e^{- t / T_p}$  & $d_0 \mathcal{D}_0 e^{- t / T_d}$  & $f_0 \mathcal{F}_0 e^{- t / T_f}$   \\
\end{tabular}
\end{center}
\label{default}
\label{Multipole}
\end{table}

To experimentally visualize the $\mathrm{V_{Si}}$ spin qudit modes, we use optically detected magnetic resonance (ODMR) as described elsewhere \cite{Kraus:2013vf,Christle:2014ti,Widmann:2014ve,Falk:2013jq,Carter:2015vc, Simin:2017iw} (see also Supplementary Methods and Fig.~S2). Optical pumping results in a preferential population of either the $m_S = \pm 3/2$ or $m_S = \pm 1/2$ states (depending on SiC polytype and the $\mathrm{V_{Si}}$ crystallographic site). Following table~\ref{Multipole}, only the spin quadrupole $\mathcal{D}_0$ is excited under these conditions, with either $d_0 > 0$ or $d_0 < 0$, respectively. Application of a strong microwave (MW) field at a fixed frequency $\nu_{\mathrm{pump}}$ mixes the $m_S = \pm 3/2$ and $m_S = \pm 1/2$ states, resulting in the excitation of other qudit modes as well. They are probed as relative difference of the spin-dependent photolumenescence $\mathrm{\Delta PL / PL}$  while sweeping the frequency of a second (weak) MW field $\nu_{\mathrm{probe}}$.

\begin{figure}[t]
\includegraphics[width=.48\textwidth]{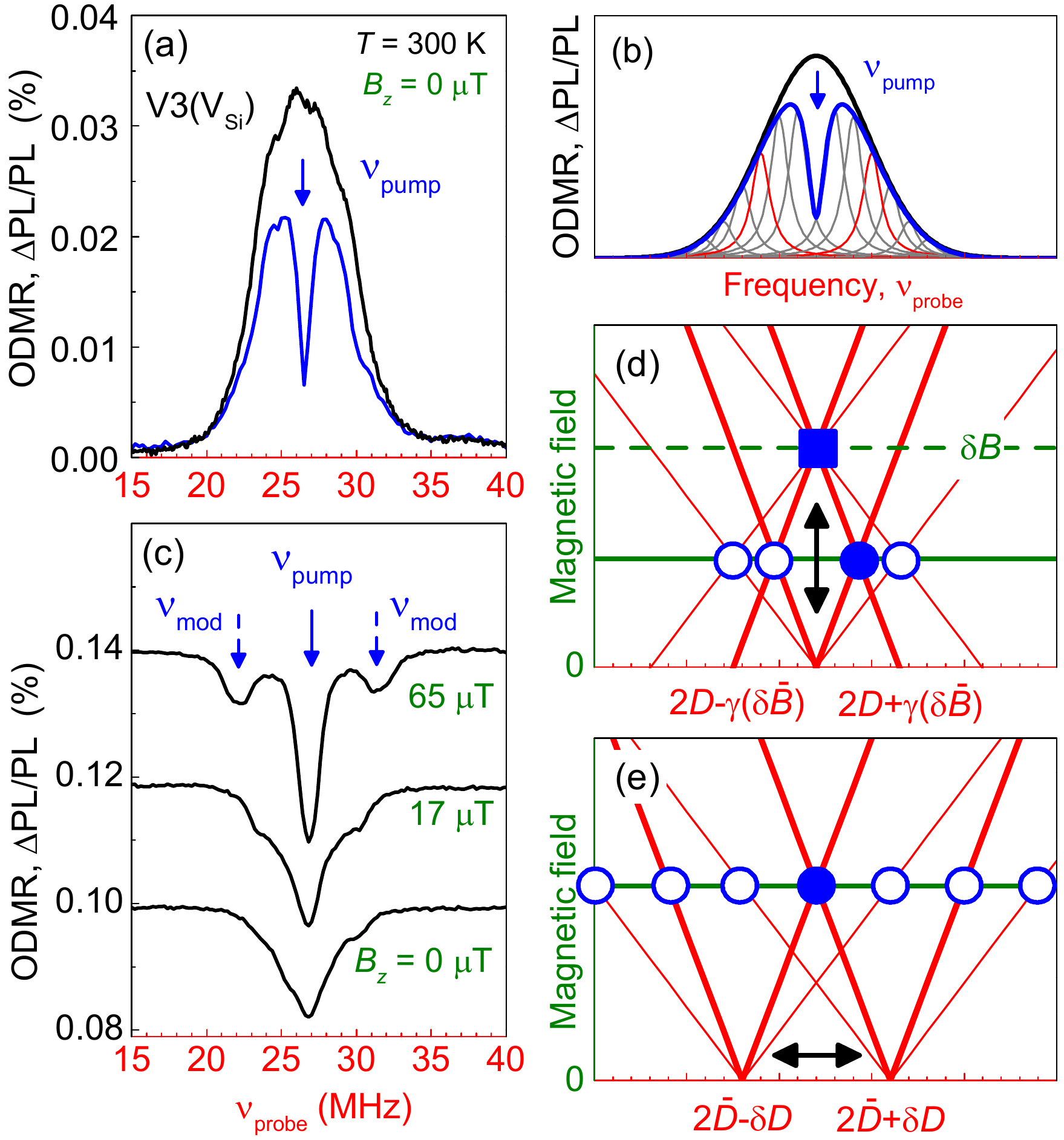}
\caption{Two-frequency ODMR spectroscopy. (a) ODMR spectrum of isotopically purified 6H-SiC without and with the pump MW field at $\nu_{\mathrm{pump}} = 26.8 \, \mathrm{MHz}$. The pump and probe powers are $W_{\mathrm{probe}} = 7 \, \mathrm{dBm}$ and $W_{\mathrm{pump}} = 14 \, \mathrm{dBm}$, respectively. (b) A schematic modelling of (a) by subtraction the contribution of a homogeneously broadened spin packet from  the Gauss-shaped ODMR line. (c) Pump-induced changes in the ODMR spectra in different magnetic fields.  The dashed arrows indicate the qudit mode frequencies. (d) and (e) The fan charts of the qudit modes as a function of the magnetic field  in case when the inhomogeneous broadening is caused by magnetic fluctuations and variation for the zero field splitting, respectively. The solid symbols correspond to the pump frequencies and the open circles correspond to the excited qudit mode frequencies. The red lines of the fan charts indicate the spectrally selected spin packets. } \label{fig1}
\end{figure}

Figure~\ref{fig1}(a) shows an ODMR resonance associated with the V3($\mathrm{V_{Si}}$) center in 6H-$^{28}$SiC \cite{Sorman:2000ij,Soltamov:2012ey,Riedel:2012jq}. The ODMR resonance without the pump MW field (black curve) reveals the quadrupole spin splitting $2D = 26.8$~MHz between the $\pm 1/2$ and $\pm 3/2$ states. Application of a strong, pump MW field at $\nu_{\mathrm{pump}} = 26.8 \, \mathrm{MHz}$ saturates the spin transitions, which is seen as a spectral hole burning \cite{Kehayias:2014dd}. To model it, we assume that the ODMR resonance is inhomogeneously broadened and agregates many homogeneous spin packets with different resonance frequencies, as schematically shown by the thin lines in Fig.~\ref{fig1}(b). The pump MW field excites qudit modes in particular spin packets, which are detected as a reduction of the ODMR signal at certain frequencies. In zero magnetic field, all the excited modes are degenerate and manifest themselves as a single spectral hole. Remarkably, the hole spectral width can be much narrower than the inhomogeneous linewidth and in certain cases we observe $250 \, \mathrm{kHz}$ (Supplementary Figs.~S3~and~S4).  

To increase the sensitivity, we modify the detection scheme. The MW pump is now modulated on/off and the ODMR spectrum is detected using a lock-in amplifier. The pump-induced changes of the ODMR spectrum are presented in Fig.~\ref{fig1}(c) for several magnetic fields $B_z$ applied along the $c$-axis of SiC. Several qudit modes $\nu_{\mathrm{mod}}$ are now clearly detected and their spectral positions relative to $\nu_{\mathrm{pump}}$ depend on $B_z$. 

To understand the spectrum of spin qudit modes, we first consider the inhomogeneous broadening caused by magnetic fluctuations (for instance, due to nuclear fields). This mechanism was shown to sufficiently influence the inhomogeneous broadening of the nitrogen-vacancy (NV) defects in diamond \cite{Kehayias:2014dd}. Upon application of the magnetic field along the $c$-axis, a single spin packet is characterized by four resonances shifting linearly with the magnetic filed as $2D \pm \gamma B_z$ and $2D \pm 2 \gamma B_z$  \cite{Simin:2016cp}. Here, $\gamma =  28 \, \mathrm{MHz / mT}$ is the gyromagnetic ratio. The corresponding fan chart is shown by the red lines in Fig.~\ref{fig1}(d). The line thickness represents the coupling strength to the MW field. Another spin packet feels another local magnetic field $\delta B$ and the fan chart is shifted along the vertical axis in Fig.~\ref{fig1}(d). The mean value $\delta \bar B$ defines the inhomogeneous ODMR linewidth. 

Assume that the pump MW field is in resonance with one of the spin transitions of a particular spin packet (the solid circle in Fig.~\ref{fig1}(d)). Then the other three spin resonances of the same spin packet will be effected due to the excitation and relaxation of spin multipoles, leading to the appearance of qudit modes  (the open circles in Fig.~\ref{fig1}(d)). Their spectral position relative to the pump frequency is symmetric in respect to the zero field splitting $2 D$. In case when $\nu_{\mathrm{pump}} = 2 D $, the excited qudit modes remain degenerate even when $B_z \neq 0$ (the square in Fig.~\ref{fig1}(d)), which contradicts to the experimental observation of Fig.~\ref{fig1}(c).

We now consider another inhomogeneous broadening mechanism, caused by the variation of the zero field splitting around the mean value $2\bar D = 26.8 \, \mathrm{MHz}$. This mechanism is natural to expect, given that we investigate isotopically-purified  6H-$^{28}$SiC with low content of the spin-carrying $^{29}$Si isotopes. In this case, the fan chart of the field-dependent ODMR lines associated with different spin packets are shifted along the horizontal axis in Fig.~\ref{fig1}(e). The positions of the excited qudit modes are given by $\nu_{\mathrm{mod}} = \nu_{\mathrm{pump}} \pm s \gamma B_z$ with $s = 1,2,3,4$. They depend on the magnetic field strength, which is in qualitative agreement with the experiment of Fig.~\ref{fig1}(c). 

\begin{figure}[t]
\includegraphics[width=.47\textwidth]{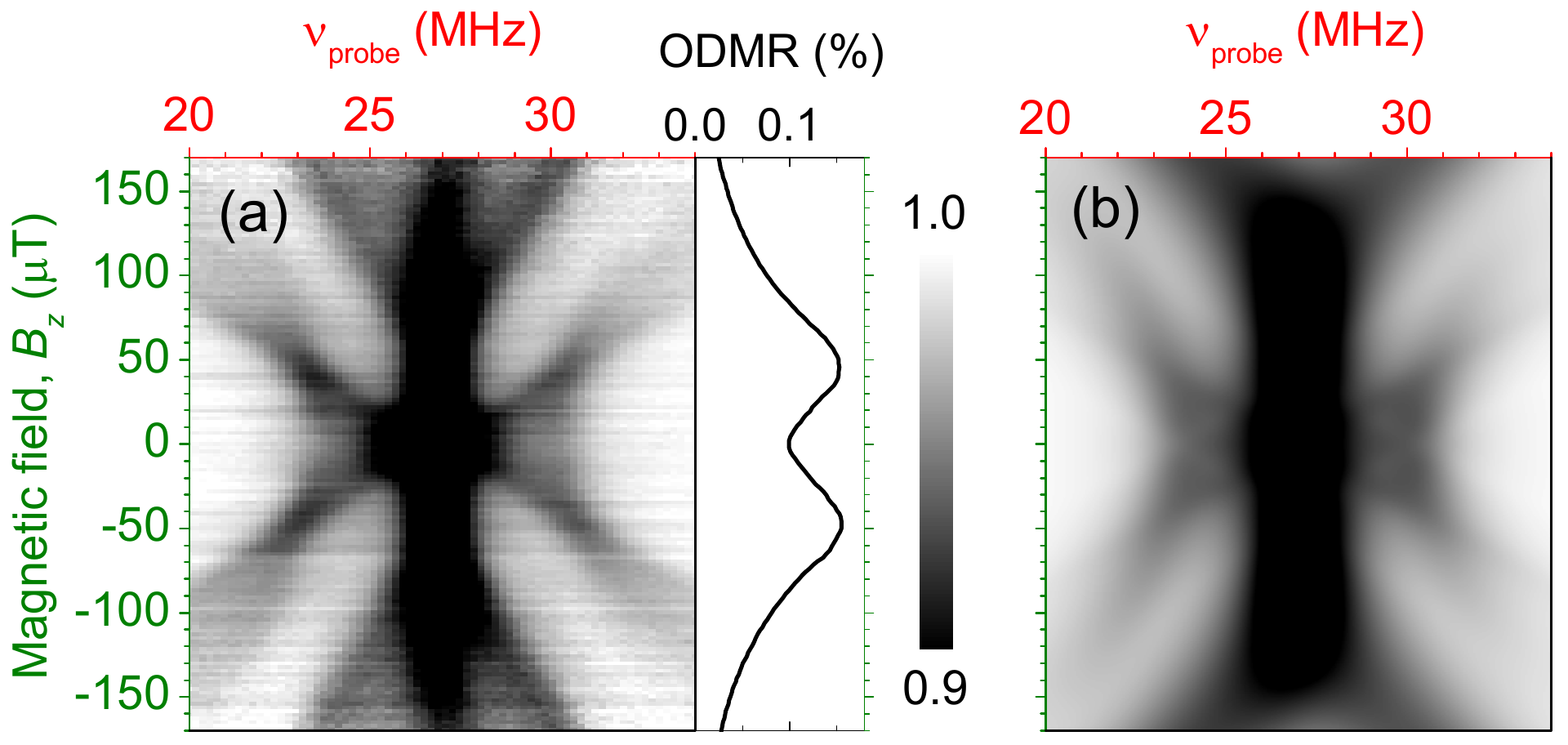}
\caption{Visualization of the excited qudit modes. (a) Magnetic field evolution of the qudit modes in an inhomogeneously broadened  spin ensemble. The ODMR signal is normalized for each magnetic field. The base ODMR signal is presented on the right side. (b) Calculated evolution of the spin qudit modes assuming $T_p = 3 T_d = 6 T_f $ (see text for details). } \label{fig2}
\end{figure}

Figure~\ref{fig2}(a) visualizes the magnetic field evolution of the V3($\mathrm{V_{Si}}$) spin qudit modes in 6H-$^{28}$SiC. The observed behavior is more complex compared to the qualitative consideration above.  Not all qudit modes shift linearly with $B_z$, they have different strength and some of them are not even detected. To understand this behavior, we have developed a theory of qudit mode excitation in inhomogeneously broadened spin ensembles.  

We start from the effective spin Hamiltonian of the $\mathrm{V_{Si}}$ centers in a simple axial model $H = D ( S_z^2 - 5/4) + \gamma \mathbf{S}  \mathbf{B}$.  In small magnetic fields that we consider, $\gamma B \ll D$, the eigenstates are given by 
\begin{align} \label{levels}
\begin{split}
&E_{``\pm3/2"} = +D \pm \frac32 \gamma B_z \,,\\
&E_{``\pm1/2"} = -D \pm \frac12 \gamma \sqrt{B_z^2 + 4 B_\perp^2} \,,
\end{split}
\end{align}
where $B_z$ and $B_\perp$ are the magnetic field components parallel and perpendicular to the $c$-axis, respectively. 
Due to the mixing of the spin states by $B_{\perp}$ and a low trigonal pyramidal local symmetry of the $\mathrm{V_{Si}}$ centers in SiC, all four transitions between the $``\pm 1/2"$ and $``\pm 3/2"$ spin states are allowed \cite{Simin:2016cp}. 
Their spectral positions read 
\begin{align}\label{satfreq}
\nu_{\rm mod}^{(s,s')} = \nu_{\rm pump} + 3 s \gamma B_z + s' \gamma \sqrt{B_z^2+4 B_\perp^2} \,,
\end{align}
where $s,s' = 0,\pm 1$. Equation~\eqref{satfreq} describes the spectral hole burning at $\nu_{\rm pump}$ ($s,s'=0$) and  8 qudit modes. 

We now analyze the strength of the qudit modes coupling to the MW field and  how excitations between these modes are transfered. At room temperature, when the energy of thermally excited phonons is much higher than the spin splitting, all spin multipoles of table~\ref{Multipole} are involved into the spin relaxation process. If spin relaxation occurs due to fluctuating magnetic fields, only the transitions with $\Delta m_S = \pm 1$ are allowed, as was also assumed in the earlier works \cite{Widmann:2014ve}. Application of this constraint to the spin relaxation matrix (Supplementary Eq.~(S14)) yields the ratio between the multipole relaxation times $T_p = 3 T_d = 6 T_f $. Remarkably, the spin relaxation rates between the $+ 3/2 \leftrightarrow + 1/2$ and $- 3/2 \leftrightarrow -1/2$ states are not equal to that  between the $+1/2 \leftrightarrow -1/2$ states, and we obtain for them $(1/2)T_d^{-1}$ and $(2/3)T_d^{-1}$, respectively. 

Following the analytical solution of the rate equations (Supplementary Eq.~(S11)), the strength of the modes $(s,s')=(+1,+1)$, $(+1,-1)$, $(-1,+1)$ and $(-1,-1)$ is proportional to $5T_d -T_p -4T_f$. Therefore, they vanish in the case if the aforementioned ratio $T_p = 3 T_d = 6 T_f $ holds. As a consequence, only 4 out of 8 possible qudit modes are observed in Fig.~\ref{fig2}(a). The calculated evolution of the qudit modes with $B_z$  is shown in Fig.~\ref{fig2}(b) (see Supplementary Theory). Here, we assume the Gaussian distribution of the zero-field splitting, $f(D) \propto \exp[(D-\bar D)^2/(\,\delta D^2)]$, with the mean value $2\bar D = 26.8 \, \mathrm{MHz}$ corresponding to the V3$(\mathrm{V_{Si}}$) center  in 6H-SiC. 
A perfect agreement with the experimental data is achieved for $B_\perp = 60\,\mu$T, accounting for a uncompensated perpendicular component of the external magnetic field. 

\begin{figure}[t]
\includegraphics[width=.47\textwidth]{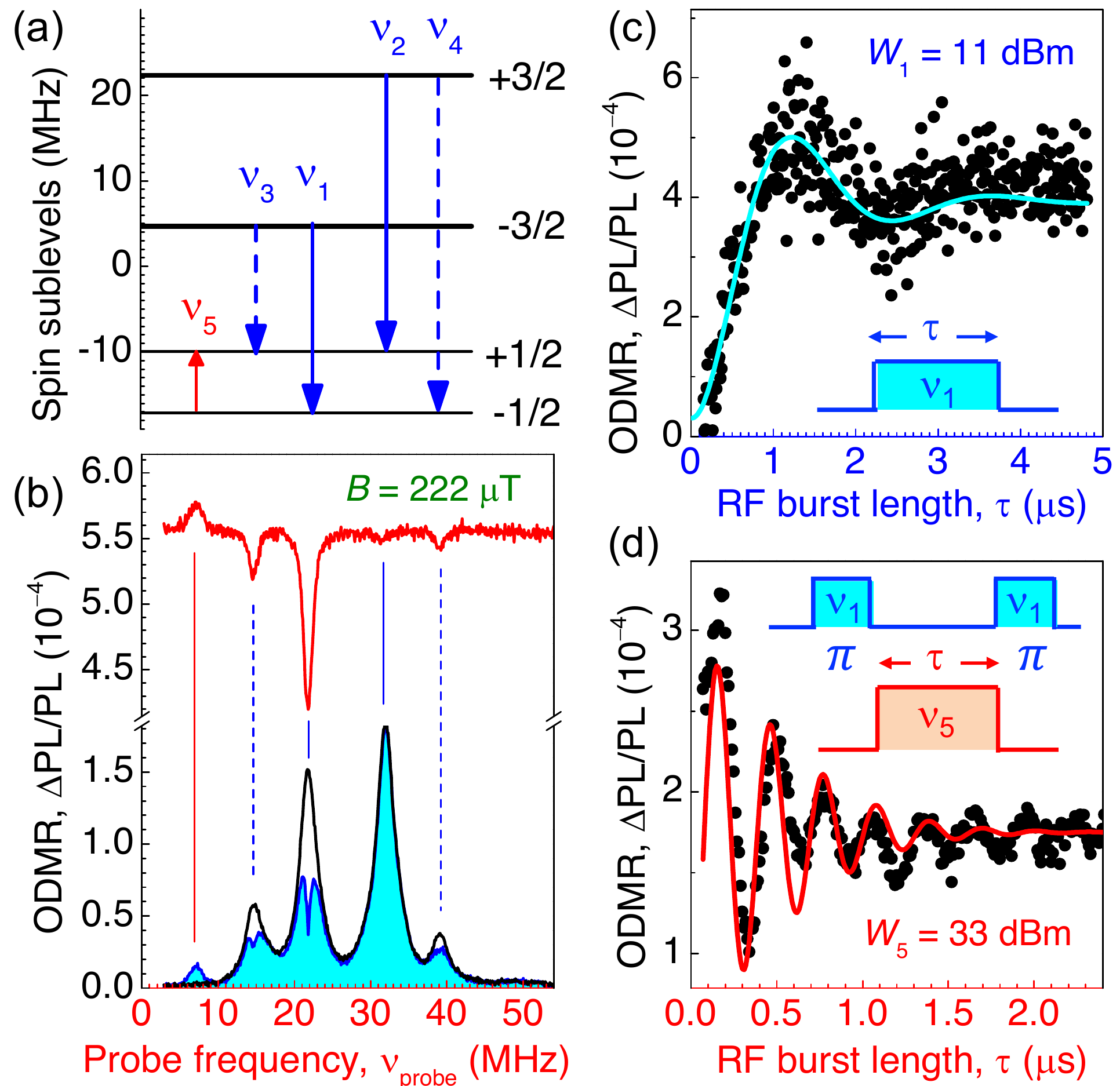}
\caption{Coherent manipulation of spin qudit modes. (a) MW-induced transitions between different spin sublevels after optical pumping into the $m_S = \pm 3/2$ states. The solid and dashed arrows correspond to the transitions with large and small matrix elements, respectively. (b) ODMR spectra in a magnetic field of $222 \,  \mathrm{\mu T}$ with and without MW pump at  $\nu_{\mathrm{pump}} = 21.8 \, \mathrm{MHz}$. The upper curve shows the spectrum of the excited qudit modes.  (c) Rabi oscillations at the $\nu_1$ resonance driven by the MW power $W_1 = 11 \,  \mathrm{dBm}$ with the corresponding $\pi$-pulse duration of $1.2 \, \mathrm{\mu s}$. (d) Rabi oscillations of a spin packet at the $\nu_5$ resonance driven by the MW power $W_5 = 33 \,  \mathrm{dBm}$.  } \label{fig3}
\end{figure}

To further confirm the conclusions of our model, we investigate the qudit modes in a stronger magnetic field $B_z = 211 \, \mathrm{\mu T}$ when the Zeeman splitting is larger than inhomogeneous broadening. A precise analysis  of the ODMR spectra \cite{Simin:2015dn, Niethammer:2016bc} shows that  $B_{\bot} = 73 \, \mathrm{\mu T}$ remains almost the same. Figure~\ref{fig3}(a) shows relative positions of the  $\mathrm{V_{Si}}$ spin sublevels calculated for this field configuration. The ODMR spectrum consists of four lines [the lower curve in Fig.~\ref{fig3}(b)] and the corresponding spin transitions enumerated from $\nu_1$ to $\nu_4$ are shown by the arrows in Fig.~\ref{fig3}(a).  The inner transitions $\nu_{1,2}$ are stronger than the outer transitions $\nu_{3,4}$, in accordance with earlier studies \cite{Simin:2016cp}. 

Now, we burn a spectrally narrow hole at $\nu_{1}$ ($\nu_{\mathrm{pump}} = 21.8 \, \mathrm{MHz}$), as shown in Fig.~\ref{fig3}(b). As expected from our model, excitation of qudit modes leads to the emergence of spectral holes at the $\nu_{2}$, $\nu_{3}$  and $\nu_{4}$ transitions [see the upper curve in Fig.~\ref{fig3}(b)]. Their spectral positions shift linearly with $\nu_{\mathrm{pump}}$, keeping the frequency difference the same. This is particularly pronounced in another sample with much larger inhomogeneous broadening, isotopically purified 4H-$^{28}$SiC (Supplementary Fig.~S5). We hence conclude that by varying $\nu_{\mathrm{pump}}$ within the $\nu_{1}$ ODMR line, different spin packets are selected. Remarkably, there is another signal at $\nu_{5}$ in Fig.~\ref{fig3}(b), which has the opposite sign and appears only if the $\nu_{1}$ qudit mode is excited. The same behavior is also observed in 4H-SiC \cite{Niethammer:2016bc} (see also Supplementary Fig.~S6). These general properties are used to implement spectrally selective coherent control of qudit modes. 

We start with driving the $\nu_{1}$ transition. Figure~\ref{fig3}(c) shows Rabi oscillations when the driving power is relatively low $W_1 = 11 \,  \mathrm{dBm}$. The corresponding $\pi$-pulse duration of $1.2 \,  \mathrm{\mu s}$ provides sub-MHz spectral selectivity. We then spectrally select a spin packet using a long $\pi$ pulse at $\nu_{1}$ and coherently drive it at the $\nu_{5}$ resonance followed by a second long $\pi$ pulse at $\nu_{1}$ for the readout [inset of Fig.~\ref{fig3}(d)]. A high driving power $W_5 = 33 \,  \mathrm{dBm}$ yields fast Rabi oscillations as presented in Fig.~\ref{fig3}(d). For such a power, the $\pi / 2$ pulse is $80 \,  \mathrm{ns}$, corresponding to a bandwidth of approximately $10 \, \mathrm{MHz}$. This bandwidth is wide
enough to encompass the $\nu_{5}$ linewidth. 

\begin{figure}[t]
\includegraphics[width=.47\textwidth]{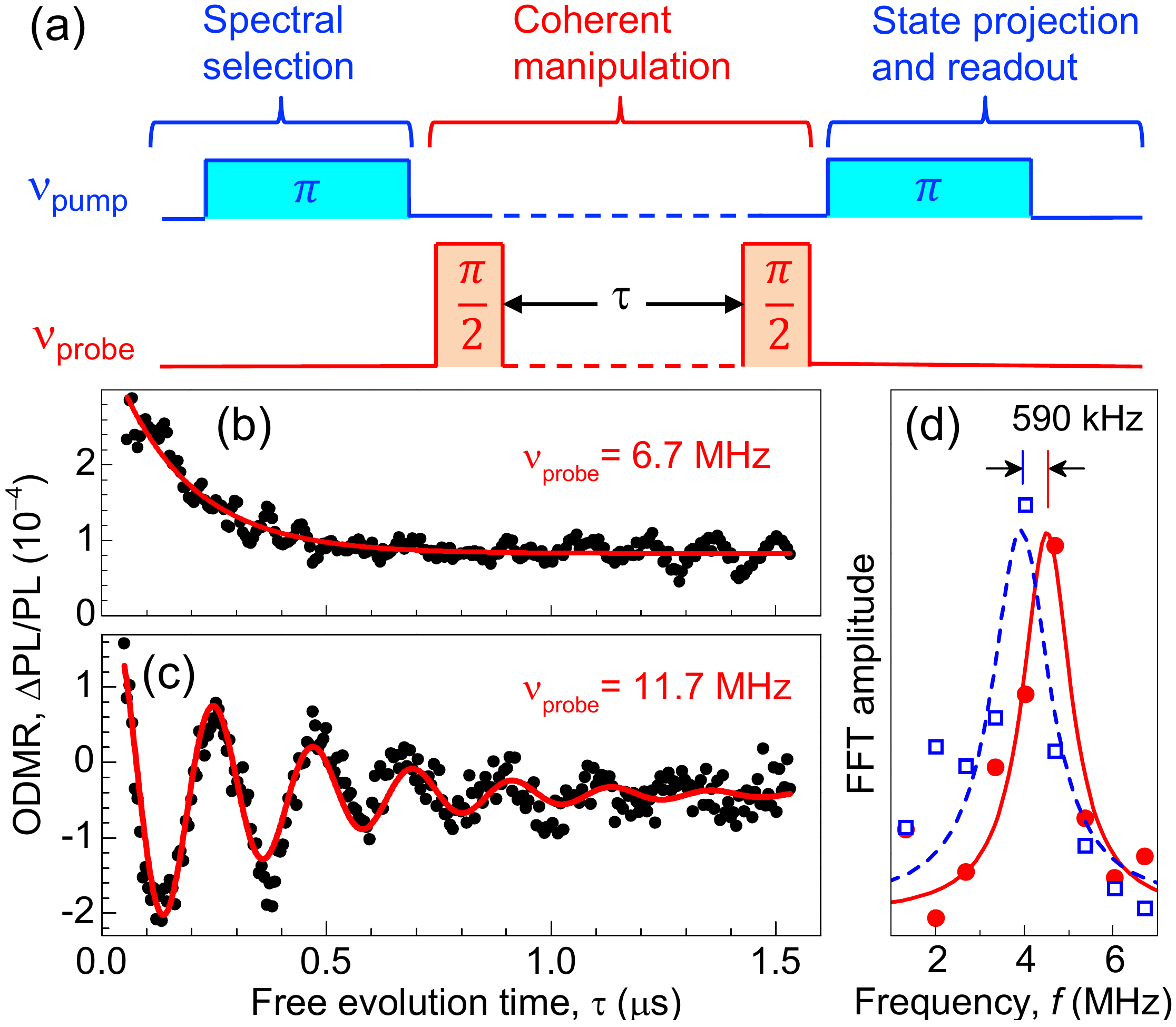}
\caption{Ramsey interferometry of spin qudit modes. (a) Pulse pattern for the spectral selection, coherent manipulation and state projection followed by the readout of spin packets. (b) Ramsey measurement of the spin packet selected by $\nu_{\mathrm{pump}} = 21.8 \, \mathrm{MHz}$ in a magnetic field of $223 \pm 1 \,  \mathrm{\mu T}$. The probe frequency $\nu_{\mathrm{probe}} = 6.7 \, \mathrm{MHz}$ is set to the $\nu_5$ resonance. The solid line represents the fit to an exponential decay with $T_2^* = 168 \pm 7 \, \mathrm{ns}$.  (c) The same as (b) but the probe frequency $\nu_{\mathrm{probe}} = 11.7 \, \mathrm{MHz}$ is detuned from the $\nu_5$ resonance. The solid line represents the fit to an exponentially decaying sinusoid with $T_2^* = 357 \pm 24 \, \mathrm{ns}$.  (d) FFT of the Ramsey fringes fitted to a Lorentz function. The solid line corresponds to the data from (c). The dashed line is for the same spin packet as in (c) but probed in a magnetic field of $242 \pm 2 \,  \mathrm{\mu T}$.  } \label{fig4}
\end{figure}

To demonstrate advantages of the qudit modes for quantum information processing and sensing, we perform two-frequency Ramsey experiments using the protocol presented in  Fig.~\ref{fig4}(a) (and Supplementary Fig.~S2).  When the probe frequency $\nu_{\mathrm{probe}} = 6.7 \, \mathrm{MHz}$ is  equal to the $\nu_{5}$ qudit mode frequency, the signal represents free induction decay [Fig.~\ref{fig4}(b)]. We note that though the other, $\nu_3$ resonance lies within the bandwidth, the population of the $m_S = -3/2$ and $m_S = +1/2$ is equal after the $\nu_1$ $\pi$-pump pulse [Fig.~\ref{fig3}(a)] and this resonance is not driven. This is also confirmed by the absence of pronounced fringes in the free induction decay of Fig.~\ref{fig4}(b). 

 When $\nu_{\mathrm{probe}} = 11.7 \, \mathrm{MHz}$ is detuned  from the $\nu_{5}$ mode, Ramsey fringes are clearly detected [Fig.~\ref{fig4}(c)].  A fit  of these dynamics to $\cos (2 \pi f_R \tau) \exp(- \tau / T_2^*)$ gives the coherence time of spin packets $T_2^* = 357 \pm 24 \, \mathrm{ns}$. For comparison, we also perform standard Ramsey measurements using single MW frequency, but do not observed fringes because of inhomogeneous broadening and non-selective excitation (Supplementary Fig.~S7). 

Solid circles in Fig.~\ref{fig4}(d) represent the fast Fourier transform (FFT) of the experimental data from Fig.~\ref{fig4}(c). The fitting of these data to a Lorentz function yields the frequency of the Ramsey fringes $f_R = 4.51 \pm 0.03 \, \mathrm{MHz}$, with the corresponding spectral resolution of $30 \, \mathrm{kHz}$. Following Eq.~(\ref{levels}), the effective magnetic field can be measured with high accuracy as 
\begin{equation}
B_{\mathrm{eff}} =  \frac{\nu_{\mathrm{probe}} - f_{\mathrm{R}}}{\gamma \sqrt{1 + 3  \sin^2 (\theta) }}  \,.
\label{Beff}
\end{equation}
With $\theta = 19^{\circ}$  from the data of Fig.~\ref{fig3}(a,b), we obtain $B_{\mathrm{eff}} = 223 \pm 1 \,  \mathrm{\mu T}$. To analyze the spectral selectivity of different spin packets, we marginally vary the external magnetitic field strength and the pump/probe frequencies. The results are summarized in Table~\ref{Ramsey-6H}, where measurement No.~1 corresponds to the case discussed above. We then increase $B$ and change $\nu_{\mathrm{pump}}$  to select the same spin packet (measurement No.~2 in Table~\ref{Ramsey-6H}). The corresponding FFT of the Ramsey fringes is represented by open squares in Fig.~\ref{fig4}(d) and a lorentzian fit (the dashed line) yields $f_R = 3.92 \pm 0.05 \, \mathrm{MHz}$ and $B_{\mathrm{eff}} = 242  \pm 2 \, \mathrm{\mu T}$. A comparison of measurements No.~1 and No.~2 demonstrates the spectral selectivity of spin packets to be approximately $600 \, \mathrm{kHz}$. We note that a unusually narrow ODMR linewidth of $500 \, \mathrm{kHz}$ in SiC  was observed earlier in two-frequency experiments, but its origin has not been discussed \cite{Niethammer:2016bc}.  

\begin{table}[tbh]
\caption{FFT frequency of the Ramsey fringes $f_{\mathrm{R}}$ in 6H-$^{28}$SiC for different pump $\nu_{\mathrm{pump}}$ and probe $\nu_{\mathrm{probe}}$ frequencies in different magnetic fields $B$ (bias coil currents $I$). Experimental data are presented in Fig.~\ref{fig4} and Supplementary Fig.~S8.}
\begin{center}
\begin{tabular}{|c|c|c|c|c|c|}
No. &  $I$ & $\nu_{\mathrm{pump}}$ & $\nu_{\mathrm{probe}}$ & $f_{\mathrm{R}}$ & $B_{\mathrm{eff}}$  \\
      &   $\mathrm{(mA)}$ & $\mathrm{(MHz)}$ & $\mathrm{(MHz)}$ &  $\mathrm{(MHz)}$ &  $\mathrm{(\mu T)}$\\
\hline
$1$ & $100$ & $21.80$ & $11.70$  & $4.51 \pm 0.03$  & $223 \pm 1$ \\
$2$ & $105$ & $21.20$ & $11.70$  & $3.92 \pm 0.05$  & $242 \pm 2$ \\
$3$ & $100$ & $21.20$ & $11.70$  & $4.51 \pm 0.04$  & $223 \pm 1$ \\
\end{tabular}
\end{center}
\label{default}
\label{Ramsey-6H}
\end{table}

Measurements No.~1 and No.~3 in Table~\ref{Ramsey-6H} demonstrate that the effective magnetic field $B_{\mathrm{eff}}$ seen by different spin packets, which are selected by different $\nu_{\mathrm{pump}}$, is the same within the error bars.  This is a manifestation that the inhomogeneous broadening is caused mostly by local variations of the zero-field splitting $2 D$ rather than magnetic fluctuations. Remarkably, onother paar of measurements (No.~2 and No.~3) in Table~\ref{Ramsey-6H} show that the magnetic field strength can be measured without calibration of the zero-field splitting, which can be used to implement absolute (i.e., immune to thermal noise and strain inhomogeneity) DC magnetometry \cite{Fang:2013dw} .  

To summarize, we have demonstrated coherent manipulation of spin qudit modes in isotopically-purified SiC at room temperature. We have also developed a theory describing the excitation and detection of these modes in inhomgeneously broadened systems and shown that qudits are characterized by multiple relaxation times. These findings can lead to dipole-coupled networks \cite{Falk:2013jq}, unconditional electron-nuclear spin registers  \cite{Klimov:2015bm} and spectral selection of highly-coherent individual spins \cite{Christle:2014ti, Widmann:2014ve, Fuchs:2015ii}, particularly in nanocrystals \cite{Muzha:2014th}. Our results hence open new possibilities to improve the sensitivity of quantum sensors and execute nontrivial quantum protocols in dense spin ensembles. 

\begin{acknowledgments}
This work has been supported by the German Research Foundation (DFG) under Grants DY 18/13 and AS 310/5,  by the ERA.Net RUS Plus program and the German Federal Ministry of Education and Research (BMBF) within project DIABASE, as well as by the Russian Science Foundation under Agreement \#16-42-01098 and RFBR \#16-02-00877. V.A.S. was supported by the Alexander von Humboldt-Foundation. A.V.P. also acknowledges the support by the RF President Grant SP-2912.2016.5 and the "BASIS" Foundation.
\end{acknowledgments}


\begin{thebibliography}{27}
\expandafter\ifx\csname natexlab\endcsname\relax\def\natexlab#1{#1}\fi
\expandafter\ifx\csname bibnamefont\endcsname\relax
  \def\bibnamefont#1{#1}\fi
\expandafter\ifx\csname bibfnamefont\endcsname\relax
  \def\bibfnamefont#1{#1}\fi
\expandafter\ifx\csname citenamefont\endcsname\relax
  \def\citenamefont#1{#1}\fi
\expandafter\ifx\csname url\endcsname\relax
  \def\url#1{\texttt{#1}}\fi
\expandafter\ifx\csname urlprefix\endcsname\relax\def\urlprefix{URL }\fi
\providecommand{\bibinfo}[2]{#2}
\providecommand{\eprint}[2][]{\url{#2}}

\bibitem[{\citenamefont{Awschalom et~al.}(2013)\citenamefont{Awschalom,
  Bassett, Dzurak, Hu, and Petta}}]{Awschalom:2013in}
\bibinfo{author}{\bibfnamefont{D.~D.} \bibnamefont{Awschalom}},
  \bibinfo{author}{\bibfnamefont{L.~C.} \bibnamefont{Bassett}},
  \bibinfo{author}{\bibfnamefont{A.~S.} \bibnamefont{Dzurak}},
  \bibinfo{author}{\bibfnamefont{E.~L.} \bibnamefont{Hu}}, \bibnamefont{and}
  \bibinfo{author}{\bibfnamefont{J.~R.} \bibnamefont{Petta}},
  \bibinfo{journal}{Science} \textbf{\bibinfo{volume}{339}},
  \bibinfo{pages}{1174} (\bibinfo{year}{2013}).

\bibitem[{\citenamefont{Kaszlikowski et~al.}(2003)\citenamefont{Kaszlikowski,
  Oi, Christandl, Chang, Ekert, Kwek, and Oh}}]{Kaszlikowski:2003js}
\bibinfo{author}{\bibfnamefont{D.}~\bibnamefont{Kaszlikowski}},
  \bibinfo{author}{\bibfnamefont{D.~K.~L.} \bibnamefont{Oi}},
  \bibinfo{author}{\bibfnamefont{M.}~\bibnamefont{Christandl}},
  \bibinfo{author}{\bibfnamefont{K.}~\bibnamefont{Chang}},
  \bibinfo{author}{\bibfnamefont{A.}~\bibnamefont{Ekert}},
  \bibinfo{author}{\bibfnamefont{L.~C.} \bibnamefont{Kwek}}, \bibnamefont{and}
  \bibinfo{author}{\bibfnamefont{C.~H.} \bibnamefont{Oh}},
  \bibinfo{journal}{Physical Review A} \textbf{\bibinfo{volume}{67}},
  \bibinfo{pages}{656} (\bibinfo{year}{2003}).

\bibitem[{\citenamefont{Lanyon et~al.}(2008)\citenamefont{Lanyon, Barbieri,
  Almeida, Jennewein, Ralph, Resch, Pryde, O'Brien, Gilchrist, and
  White}}]{Lanyon:2008gvc}
\bibinfo{author}{\bibfnamefont{B.~P.} \bibnamefont{Lanyon}},
  \bibinfo{author}{\bibfnamefont{M.}~\bibnamefont{Barbieri}},
  \bibinfo{author}{\bibfnamefont{M.~P.} \bibnamefont{Almeida}},
  \bibinfo{author}{\bibfnamefont{T.}~\bibnamefont{Jennewein}},
  \bibinfo{author}{\bibfnamefont{T.~C.} \bibnamefont{Ralph}},
  \bibinfo{author}{\bibfnamefont{K.~J.} \bibnamefont{Resch}},
  \bibinfo{author}{\bibfnamefont{G.~J.} \bibnamefont{Pryde}},
  \bibinfo{author}{\bibfnamefont{J.~L.} \bibnamefont{O'Brien}},
  \bibinfo{author}{\bibfnamefont{A.}~\bibnamefont{Gilchrist}},
  \bibnamefont{and} \bibinfo{author}{\bibfnamefont{A.~G.} \bibnamefont{White}},
  \bibinfo{journal}{Nature Physics} \textbf{\bibinfo{volume}{5}},
  \bibinfo{pages}{134} (\bibinfo{year}{2008}).

\bibitem[{\citenamefont{Shlyakhov et~al.}(2018)\citenamefont{Shlyakhov,
  Zemlyanov, Suslov, Lebedev, Paraoanu, Lesovik, and
  Blatter}}]{Shlyakhov:2018dg}
\bibinfo{author}{\bibfnamefont{A.~R.} \bibnamefont{Shlyakhov}},
  \bibinfo{author}{\bibfnamefont{V.~V.} \bibnamefont{Zemlyanov}},
  \bibinfo{author}{\bibfnamefont{M.~V.} \bibnamefont{Suslov}},
  \bibinfo{author}{\bibfnamefont{A.~V.} \bibnamefont{Lebedev}},
  \bibinfo{author}{\bibfnamefont{G.~S.} \bibnamefont{Paraoanu}},
  \bibinfo{author}{\bibfnamefont{G.~B.} \bibnamefont{Lesovik}},
  \bibnamefont{and} \bibinfo{author}{\bibfnamefont{G.}~\bibnamefont{Blatter}},
  \bibinfo{journal}{Physical Review A} \textbf{\bibinfo{volume}{97}},
  \bibinfo{pages}{022115} (\bibinfo{year}{2018}).

\bibitem[{\citenamefont{Kues et~al.}(2017)\citenamefont{Kues, Reimer, Roztocki,
  Cort{\'e}s, Sciara, Wetzel, Zhang, Cino, Chu, Little et~al.}}]{Kues:2017db}
\bibinfo{author}{\bibfnamefont{M.}~\bibnamefont{Kues}},
  \bibinfo{author}{\bibfnamefont{C.}~\bibnamefont{Reimer}},
  \bibinfo{author}{\bibfnamefont{P.}~\bibnamefont{Roztocki}},
  \bibinfo{author}{\bibfnamefont{L.~R.} \bibnamefont{Cort{\'e}s}},
  \bibinfo{author}{\bibfnamefont{S.}~\bibnamefont{Sciara}},
  \bibinfo{author}{\bibfnamefont{B.}~\bibnamefont{Wetzel}},
  \bibinfo{author}{\bibfnamefont{Y.}~\bibnamefont{Zhang}},
  \bibinfo{author}{\bibfnamefont{A.}~\bibnamefont{Cino}},
  \bibinfo{author}{\bibfnamefont{S.~T.} \bibnamefont{Chu}},
  \bibinfo{author}{\bibfnamefont{B.~E.} \bibnamefont{Little}},
  \bibnamefont{et~al.}, \bibinfo{journal}{Nature}
  \textbf{\bibinfo{volume}{546}}, \bibinfo{pages}{622} (\bibinfo{year}{2017}).

\bibitem[{\citenamefont{Neeley et~al.}(2009)\citenamefont{Neeley, Ansmann,
  Bialczak, Hofheinz, Lucero, O'Connell, Sank, Wang, Wenner, Cleland
  et~al.}}]{Neeley:2009ic}
\bibinfo{author}{\bibfnamefont{M.}~\bibnamefont{Neeley}},
  \bibinfo{author}{\bibfnamefont{M.}~\bibnamefont{Ansmann}},
  \bibinfo{author}{\bibfnamefont{R.~C.} \bibnamefont{Bialczak}},
  \bibinfo{author}{\bibfnamefont{M.}~\bibnamefont{Hofheinz}},
  \bibinfo{author}{\bibfnamefont{E.}~\bibnamefont{Lucero}},
  \bibinfo{author}{\bibfnamefont{A.~D.} \bibnamefont{O'Connell}},
  \bibinfo{author}{\bibfnamefont{D.}~\bibnamefont{Sank}},
  \bibinfo{author}{\bibfnamefont{H.}~\bibnamefont{Wang}},
  \bibinfo{author}{\bibfnamefont{J.}~\bibnamefont{Wenner}},
  \bibinfo{author}{\bibfnamefont{A.~N.} \bibnamefont{Cleland}},
  \bibnamefont{et~al.}, \bibinfo{journal}{Science}
  \textbf{\bibinfo{volume}{325}}, \bibinfo{pages}{722} (\bibinfo{year}{2009}).

\bibitem[{\citenamefont{Riedel et~al.}(2012)\citenamefont{Riedel, Fuchs, Kraus,
  V{\"a}th, Sperlich, Dyakonov, Soltamova, Baranov, Ilyin, and
  Astakhov}}]{Riedel:2012jq}
\bibinfo{author}{\bibfnamefont{D.}~\bibnamefont{Riedel}},
  \bibinfo{author}{\bibfnamefont{F.}~\bibnamefont{Fuchs}},
  \bibinfo{author}{\bibfnamefont{H.}~\bibnamefont{Kraus}},
  \bibinfo{author}{\bibfnamefont{S.}~\bibnamefont{V{\"a}th}},
  \bibinfo{author}{\bibfnamefont{A.}~\bibnamefont{Sperlich}},
  \bibinfo{author}{\bibfnamefont{V.}~\bibnamefont{Dyakonov}},
  \bibinfo{author}{\bibfnamefont{A.}~\bibnamefont{Soltamova}},
  \bibinfo{author}{\bibfnamefont{P.}~\bibnamefont{Baranov}},
  \bibinfo{author}{\bibfnamefont{V.}~\bibnamefont{Ilyin}}, \bibnamefont{and}
  \bibinfo{author}{\bibfnamefont{G.~V.} \bibnamefont{Astakhov}},
  \bibinfo{journal}{Physical Review Letters} \textbf{\bibinfo{volume}{109}},
  \bibinfo{pages}{226402} (\bibinfo{year}{2012}).

\bibitem[{\citenamefont{Mizuochi et~al.}(2002)\citenamefont{Mizuochi, Yamasaki,
  Takizawa, Morishita, Ohshima, Itoh, and Isoya}}]{Mizuochi:2002kl}
\bibinfo{author}{\bibfnamefont{N.}~\bibnamefont{Mizuochi}},
  \bibinfo{author}{\bibfnamefont{S.}~\bibnamefont{Yamasaki}},
  \bibinfo{author}{\bibfnamefont{H.}~\bibnamefont{Takizawa}},
  \bibinfo{author}{\bibfnamefont{N.}~\bibnamefont{Morishita}},
  \bibinfo{author}{\bibfnamefont{T.}~\bibnamefont{Ohshima}},
  \bibinfo{author}{\bibfnamefont{H.}~\bibnamefont{Itoh}}, \bibnamefont{and}
  \bibinfo{author}{\bibfnamefont{J.}~\bibnamefont{Isoya}},
  \bibinfo{journal}{Physical Review B} \textbf{\bibinfo{volume}{66}},
  \bibinfo{pages}{235202} (\bibinfo{year}{2002}).

\bibitem[{\citenamefont{Kraus et~al.}(2014{\natexlab{a}})\citenamefont{Kraus,
  Soltamov, Riedel, V{\"a}th, Fuchs, Sperlich, Baranov, Dyakonov, and
  Astakhov}}]{Kraus:2013di}
\bibinfo{author}{\bibfnamefont{H.}~\bibnamefont{Kraus}},
  \bibinfo{author}{\bibfnamefont{V.~A.} \bibnamefont{Soltamov}},
  \bibinfo{author}{\bibfnamefont{D.}~\bibnamefont{Riedel}},
  \bibinfo{author}{\bibfnamefont{S.}~\bibnamefont{V{\"a}th}},
  \bibinfo{author}{\bibfnamefont{F.}~\bibnamefont{Fuchs}},
  \bibinfo{author}{\bibfnamefont{A.}~\bibnamefont{Sperlich}},
  \bibinfo{author}{\bibfnamefont{P.~G.} \bibnamefont{Baranov}},
  \bibinfo{author}{\bibfnamefont{V.}~\bibnamefont{Dyakonov}}, \bibnamefont{and}
  \bibinfo{author}{\bibfnamefont{G.~V.} \bibnamefont{Astakhov}},
  \bibinfo{journal}{Nature Physics} \textbf{\bibinfo{volume}{10}},
  \bibinfo{pages}{157} (\bibinfo{year}{2014}{\natexlab{a}}).

\bibitem[{\citenamefont{Widmann et~al.}(2015)\citenamefont{Widmann, Lee,
  Rendler, Son, Fedder, Paik, Yang, Zhao, Yang, Booker
  et~al.}}]{Widmann:2014ve}
\bibinfo{author}{\bibfnamefont{M.}~\bibnamefont{Widmann}},
  \bibinfo{author}{\bibfnamefont{S.-Y.} \bibnamefont{Lee}},
  \bibinfo{author}{\bibfnamefont{T.}~\bibnamefont{Rendler}},
  \bibinfo{author}{\bibfnamefont{N.~T.} \bibnamefont{Son}},
  \bibinfo{author}{\bibfnamefont{H.}~\bibnamefont{Fedder}},
  \bibinfo{author}{\bibfnamefont{S.}~\bibnamefont{Paik}},
  \bibinfo{author}{\bibfnamefont{L.-P.} \bibnamefont{Yang}},
  \bibinfo{author}{\bibfnamefont{N.}~\bibnamefont{Zhao}},
  \bibinfo{author}{\bibfnamefont{S.}~\bibnamefont{Yang}},
  \bibinfo{author}{\bibfnamefont{I.}~\bibnamefont{Booker}},
  \bibnamefont{et~al.}, \bibinfo{journal}{Nature Materials}
  \textbf{\bibinfo{volume}{14}}, \bibinfo{pages}{164} (\bibinfo{year}{2015}).

\bibitem[{\citenamefont{Carter et~al.}(2015)\citenamefont{Carter, Soykal, Dev,
  Economou, and Glaser}}]{Carter:2015vc}
\bibinfo{author}{\bibfnamefont{S.~G.} \bibnamefont{Carter}},
  \bibinfo{author}{\bibfnamefont{{\"O}.~O.} \bibnamefont{Soykal}},
  \bibinfo{author}{\bibfnamefont{P.}~\bibnamefont{Dev}},
  \bibinfo{author}{\bibfnamefont{S.~E.} \bibnamefont{Economou}},
  \bibnamefont{and} \bibinfo{author}{\bibfnamefont{E.~R.}
  \bibnamefont{Glaser}}, \bibinfo{journal}{Physical Review B}
  \textbf{\bibinfo{volume}{92}}, \bibinfo{pages}{161202}
  (\bibinfo{year}{2015}).

\bibitem[{\citenamefont{Simin et~al.}(2017)\citenamefont{Simin, Kraus,
  Sperlich, Ohshima, Astakhov, and Dyakonov}}]{Simin:2017iw}
\bibinfo{author}{\bibfnamefont{D.}~\bibnamefont{Simin}},
  \bibinfo{author}{\bibfnamefont{H.}~\bibnamefont{Kraus}},
  \bibinfo{author}{\bibfnamefont{A.}~\bibnamefont{Sperlich}},
  \bibinfo{author}{\bibfnamefont{T.}~\bibnamefont{Ohshima}},
  \bibinfo{author}{\bibfnamefont{G.~V.} \bibnamefont{Astakhov}},
  \bibnamefont{and} \bibinfo{author}{\bibfnamefont{V.}~\bibnamefont{Dyakonov}},
  \bibinfo{journal}{Physical Review B} \textbf{\bibinfo{volume}{95}},
  \bibinfo{pages}{161201(R)} (\bibinfo{year}{2017}).

\bibitem[{\citenamefont{Tairov and Tsvetkov}(1978)}]{Tairov:1978hd}
\bibinfo{author}{\bibfnamefont{Y.~M.} \bibnamefont{Tairov}} \bibnamefont{and}
  \bibinfo{author}{\bibfnamefont{V.~F.} \bibnamefont{Tsvetkov}},
  \bibinfo{journal}{Journal of Crystal Growth} \textbf{\bibinfo{volume}{43}},
  \bibinfo{pages}{209} (\bibinfo{year}{1978}).

\bibitem[{\citenamefont{Simin et~al.}(2016)\citenamefont{Simin, Soltamov,
  Poshakinskiy, Anisimov, Babunts, Tolmachev, Mokhov, Trupke, Tarasenko,
  Sperlich et~al.}}]{Simin:2016cp}
\bibinfo{author}{\bibfnamefont{D.}~\bibnamefont{Simin}},
  \bibinfo{author}{\bibfnamefont{V.~A.} \bibnamefont{Soltamov}},
  \bibinfo{author}{\bibfnamefont{A.~V.} \bibnamefont{Poshakinskiy}},
  \bibinfo{author}{\bibfnamefont{A.~N.} \bibnamefont{Anisimov}},
  \bibinfo{author}{\bibfnamefont{R.~A.} \bibnamefont{Babunts}},
  \bibinfo{author}{\bibfnamefont{D.~O.} \bibnamefont{Tolmachev}},
  \bibinfo{author}{\bibfnamefont{E.~N.} \bibnamefont{Mokhov}},
  \bibinfo{author}{\bibfnamefont{M.}~\bibnamefont{Trupke}},
  \bibinfo{author}{\bibfnamefont{S.~A.} \bibnamefont{Tarasenko}},
  \bibinfo{author}{\bibfnamefont{A.}~\bibnamefont{Sperlich}},
  \bibnamefont{et~al.}, \bibinfo{journal}{Physical Review X}
  \textbf{\bibinfo{volume}{6}}, \bibinfo{pages}{031014} (\bibinfo{year}{2016}).

\bibitem[{\citenamefont{Tarasenko et~al.}(2018)\citenamefont{Tarasenko,
  Poshakinskiy, Simin, Soltamov, Mokhov, Baranov, Dyakonov, and
  Astakhov}}]{Tarasenko:2017ky}
\bibinfo{author}{\bibfnamefont{S.~A.} \bibnamefont{Tarasenko}},
  \bibinfo{author}{\bibfnamefont{A.~V.} \bibnamefont{Poshakinskiy}},
  \bibinfo{author}{\bibfnamefont{D.}~\bibnamefont{Simin}},
  \bibinfo{author}{\bibfnamefont{V.~A.} \bibnamefont{Soltamov}},
  \bibinfo{author}{\bibfnamefont{E.~N.} \bibnamefont{Mokhov}},
  \bibinfo{author}{\bibfnamefont{P.~G.} \bibnamefont{Baranov}},
  \bibinfo{author}{\bibfnamefont{V.}~\bibnamefont{Dyakonov}}, \bibnamefont{and}
  \bibinfo{author}{\bibfnamefont{G.~V.} \bibnamefont{Astakhov}},
  \bibinfo{journal}{physica status solidi (b)} \textbf{\bibinfo{volume}{255}},
  \bibinfo{pages}{1700258} (\bibinfo{year}{2018}).

\bibitem[{\citenamefont{Kraus et~al.}(2014{\natexlab{b}})\citenamefont{Kraus,
  Soltamov, Fuchs, Simin, Sperlich, Baranov, Astakhov, and
  Dyakonov}}]{Kraus:2013vf}
\bibinfo{author}{\bibfnamefont{H.}~\bibnamefont{Kraus}},
  \bibinfo{author}{\bibfnamefont{V.~A.} \bibnamefont{Soltamov}},
  \bibinfo{author}{\bibfnamefont{F.}~\bibnamefont{Fuchs}},
  \bibinfo{author}{\bibfnamefont{D.}~\bibnamefont{Simin}},
  \bibinfo{author}{\bibfnamefont{A.}~\bibnamefont{Sperlich}},
  \bibinfo{author}{\bibfnamefont{P.~G.} \bibnamefont{Baranov}},
  \bibinfo{author}{\bibfnamefont{G.~V.} \bibnamefont{Astakhov}},
  \bibnamefont{and} \bibinfo{author}{\bibfnamefont{V.}~\bibnamefont{Dyakonov}},
  \bibinfo{journal}{Scientific Reports} \textbf{\bibinfo{volume}{4}},
  \bibinfo{pages}{5303} (\bibinfo{year}{2014}{\natexlab{b}}).

\bibitem[{\citenamefont{Christle et~al.}(2015)\citenamefont{Christle, Falk,
  Andrich, Klimov, Hassan, Son, Janz{\'e}n, Ohshima, and
  Awschalom}}]{Christle:2014ti}
\bibinfo{author}{\bibfnamefont{D.~J.} \bibnamefont{Christle}},
  \bibinfo{author}{\bibfnamefont{A.~L.} \bibnamefont{Falk}},
  \bibinfo{author}{\bibfnamefont{P.}~\bibnamefont{Andrich}},
  \bibinfo{author}{\bibfnamefont{P.~V.} \bibnamefont{Klimov}},
  \bibinfo{author}{\bibfnamefont{J.~u.} \bibnamefont{Hassan}},
  \bibinfo{author}{\bibfnamefont{N.~T.} \bibnamefont{Son}},
  \bibinfo{author}{\bibfnamefont{E.}~\bibnamefont{Janz{\'e}n}},
  \bibinfo{author}{\bibfnamefont{T.}~\bibnamefont{Ohshima}}, \bibnamefont{and}
  \bibinfo{author}{\bibfnamefont{D.~D.} \bibnamefont{Awschalom}},
  \bibinfo{journal}{Nature Materials} \textbf{\bibinfo{volume}{14}},
  \bibinfo{pages}{160} (\bibinfo{year}{2015}).

\bibitem[{\citenamefont{Falk et~al.}(2013)\citenamefont{Falk, Buckley,
  Calusine, Koehl, Dobrovitski, Politi, Zorman, Feng, and
  Awschalom}}]{Falk:2013jq}
\bibinfo{author}{\bibfnamefont{A.~L.} \bibnamefont{Falk}},
  \bibinfo{author}{\bibfnamefont{B.~B.} \bibnamefont{Buckley}},
  \bibinfo{author}{\bibfnamefont{G.}~\bibnamefont{Calusine}},
  \bibinfo{author}{\bibfnamefont{W.~F.} \bibnamefont{Koehl}},
  \bibinfo{author}{\bibfnamefont{V.~V.} \bibnamefont{Dobrovitski}},
  \bibinfo{author}{\bibfnamefont{A.}~\bibnamefont{Politi}},
  \bibinfo{author}{\bibfnamefont{C.~A.} \bibnamefont{Zorman}},
  \bibinfo{author}{\bibfnamefont{P.~X.~L.} \bibnamefont{Feng}},
  \bibnamefont{and} \bibinfo{author}{\bibfnamefont{D.~D.}
  \bibnamefont{Awschalom}}, \bibinfo{journal}{Nature Communications}
  \textbf{\bibinfo{volume}{4}}, \bibinfo{pages}{1819} (\bibinfo{year}{2013}).

\bibitem[{\citenamefont{S{\"o}rman et~al.}(2000)\citenamefont{S{\"o}rman, Son,
  Chen, Kordina, Hallin, and Janz{\'e}n}}]{Sorman:2000ij}
\bibinfo{author}{\bibfnamefont{E.}~\bibnamefont{S{\"o}rman}},
  \bibinfo{author}{\bibfnamefont{N.}~\bibnamefont{Son}},
  \bibinfo{author}{\bibfnamefont{W.}~\bibnamefont{Chen}},
  \bibinfo{author}{\bibfnamefont{O.}~\bibnamefont{Kordina}},
  \bibinfo{author}{\bibfnamefont{C.}~\bibnamefont{Hallin}}, \bibnamefont{and}
  \bibinfo{author}{\bibfnamefont{E.}~\bibnamefont{Janz{\'e}n}},
  \bibinfo{journal}{Physical Review B} \textbf{\bibinfo{volume}{61}},
  \bibinfo{pages}{2613} (\bibinfo{year}{2000}).

\bibitem[{\citenamefont{Soltamov et~al.}(2012)\citenamefont{Soltamov,
  Soltamova, Baranov, and Proskuryakov}}]{Soltamov:2012ey}
\bibinfo{author}{\bibfnamefont{V.~A.} \bibnamefont{Soltamov}},
  \bibinfo{author}{\bibfnamefont{A.~A.} \bibnamefont{Soltamova}},
  \bibinfo{author}{\bibfnamefont{P.~G.} \bibnamefont{Baranov}},
  \bibnamefont{and} \bibinfo{author}{\bibfnamefont{I.~I.}
  \bibnamefont{Proskuryakov}}, \bibinfo{journal}{Physical Review Letters}
  \textbf{\bibinfo{volume}{108}}, \bibinfo{pages}{226402}
  (\bibinfo{year}{2012}).

\bibitem[{\citenamefont{Kehayias et~al.}(2014)\citenamefont{Kehayias,
  Mr{\'o}zek, Acosta, Jarmola, Rudnicki, Folman, Gawlik, and
  Budker}}]{Kehayias:2014dd}
\bibinfo{author}{\bibfnamefont{P.}~\bibnamefont{Kehayias}},
  \bibinfo{author}{\bibfnamefont{M.}~\bibnamefont{Mr{\'o}zek}},
  \bibinfo{author}{\bibfnamefont{V.~M.} \bibnamefont{Acosta}},
  \bibinfo{author}{\bibfnamefont{A.}~\bibnamefont{Jarmola}},
  \bibinfo{author}{\bibfnamefont{D.~S.} \bibnamefont{Rudnicki}},
  \bibinfo{author}{\bibfnamefont{R.}~\bibnamefont{Folman}},
  \bibinfo{author}{\bibfnamefont{W.}~\bibnamefont{Gawlik}}, \bibnamefont{and}
  \bibinfo{author}{\bibfnamefont{D.}~\bibnamefont{Budker}},
  \bibinfo{journal}{Physical Review B} \textbf{\bibinfo{volume}{89}},
  \bibinfo{pages}{245202} (\bibinfo{year}{2014}).

\bibitem[{\citenamefont{Simin et~al.}(2015)\citenamefont{Simin, Fuchs, Kraus,
  Sperlich, Baranov, Astakhov, and Dyakonov}}]{Simin:2015dn}
\bibinfo{author}{\bibfnamefont{D.}~\bibnamefont{Simin}},
  \bibinfo{author}{\bibfnamefont{F.}~\bibnamefont{Fuchs}},
  \bibinfo{author}{\bibfnamefont{H.}~\bibnamefont{Kraus}},
  \bibinfo{author}{\bibfnamefont{A.}~\bibnamefont{Sperlich}},
  \bibinfo{author}{\bibfnamefont{P.~G.} \bibnamefont{Baranov}},
  \bibinfo{author}{\bibfnamefont{G.~V.} \bibnamefont{Astakhov}},
  \bibnamefont{and} \bibinfo{author}{\bibfnamefont{V.}~\bibnamefont{Dyakonov}},
  \bibinfo{journal}{Physical Review Applied} \textbf{\bibinfo{volume}{4}},
  \bibinfo{pages}{014009} (\bibinfo{year}{2015}).

\bibitem[{\citenamefont{Niethammer et~al.}(2016)\citenamefont{Niethammer,
  Widmann, Lee, Stenberg, Kordina, Ohshima, Son, Janz{\'e}n, and
  Wrachtrup}}]{Niethammer:2016bc}
\bibinfo{author}{\bibfnamefont{M.}~\bibnamefont{Niethammer}},
  \bibinfo{author}{\bibfnamefont{M.}~\bibnamefont{Widmann}},
  \bibinfo{author}{\bibfnamefont{S.-Y.} \bibnamefont{Lee}},
  \bibinfo{author}{\bibfnamefont{P.}~\bibnamefont{Stenberg}},
  \bibinfo{author}{\bibfnamefont{O.}~\bibnamefont{Kordina}},
  \bibinfo{author}{\bibfnamefont{T.}~\bibnamefont{Ohshima}},
  \bibinfo{author}{\bibfnamefont{N.~T.} \bibnamefont{Son}},
  \bibinfo{author}{\bibfnamefont{E.}~\bibnamefont{Janz{\'e}n}},
  \bibnamefont{and}
  \bibinfo{author}{\bibfnamefont{J.}~\bibnamefont{Wrachtrup}},
  \bibinfo{journal}{Physical Review Applied} \textbf{\bibinfo{volume}{6}},
  \bibinfo{pages}{034001} (\bibinfo{year}{2016}).

\bibitem[{\citenamefont{Fang et~al.}(2013)\citenamefont{Fang, Acosta, Santori,
  Huang, Itoh, Watanabe, Shikata, and Beausoleil}}]{Fang:2013dw}
\bibinfo{author}{\bibfnamefont{K.}~\bibnamefont{Fang}},
  \bibinfo{author}{\bibfnamefont{V.~M.} \bibnamefont{Acosta}},
  \bibinfo{author}{\bibfnamefont{C.}~\bibnamefont{Santori}},
  \bibinfo{author}{\bibfnamefont{Z.}~\bibnamefont{Huang}},
  \bibinfo{author}{\bibfnamefont{K.~M.} \bibnamefont{Itoh}},
  \bibinfo{author}{\bibfnamefont{H.}~\bibnamefont{Watanabe}},
  \bibinfo{author}{\bibfnamefont{S.}~\bibnamefont{Shikata}}, \bibnamefont{and}
  \bibinfo{author}{\bibfnamefont{R.~G.} \bibnamefont{Beausoleil}},
  \bibinfo{journal}{Physical Review Letters} \textbf{\bibinfo{volume}{110}},
  \bibinfo{pages}{130802} (\bibinfo{year}{2013}).

\bibitem[{\citenamefont{Klimov et~al.}(2015)\citenamefont{Klimov, Falk,
  Christle, Dobrovitski, and Awschalom}}]{Klimov:2015bm}
\bibinfo{author}{\bibfnamefont{P.~V.} \bibnamefont{Klimov}},
  \bibinfo{author}{\bibfnamefont{A.~L.} \bibnamefont{Falk}},
  \bibinfo{author}{\bibfnamefont{D.~J.} \bibnamefont{Christle}},
  \bibinfo{author}{\bibfnamefont{V.~V.} \bibnamefont{Dobrovitski}},
  \bibnamefont{and} \bibinfo{author}{\bibfnamefont{D.~D.}
  \bibnamefont{Awschalom}}, \bibinfo{journal}{Science Advances}
  \textbf{\bibinfo{volume}{1}}, \bibinfo{pages}{e1501015}
  (\bibinfo{year}{2015}).

\bibitem[{\citenamefont{Fuchs et~al.}(2015)\citenamefont{Fuchs, Stender,
  Trupke, Simin, Pflaum, Dyakonov, and Astakhov}}]{Fuchs:2015ii}
\bibinfo{author}{\bibfnamefont{F.}~\bibnamefont{Fuchs}},
  \bibinfo{author}{\bibfnamefont{B.}~\bibnamefont{Stender}},
  \bibinfo{author}{\bibfnamefont{M.}~\bibnamefont{Trupke}},
  \bibinfo{author}{\bibfnamefont{D.}~\bibnamefont{Simin}},
  \bibinfo{author}{\bibfnamefont{J.}~\bibnamefont{Pflaum}},
  \bibinfo{author}{\bibfnamefont{V.}~\bibnamefont{Dyakonov}}, \bibnamefont{and}
  \bibinfo{author}{\bibfnamefont{G.~V.} \bibnamefont{Astakhov}},
  \bibinfo{journal}{Nature Communications} \textbf{\bibinfo{volume}{6}},
  \bibinfo{pages}{7578} (\bibinfo{year}{2015}).

\bibitem[{\citenamefont{Muzha et~al.}(2014)\citenamefont{Muzha, Fuchs,
  Tarakina, Simin, Trupke, Soltamov, Mokhov, Baranov, Dyakonov, Krueger
  et~al.}}]{Muzha:2014th}
\bibinfo{author}{\bibfnamefont{A.}~\bibnamefont{Muzha}},
  \bibinfo{author}{\bibfnamefont{F.}~\bibnamefont{Fuchs}},
  \bibinfo{author}{\bibfnamefont{N.~V.} \bibnamefont{Tarakina}},
  \bibinfo{author}{\bibfnamefont{D.}~\bibnamefont{Simin}},
  \bibinfo{author}{\bibfnamefont{M.}~\bibnamefont{Trupke}},
  \bibinfo{author}{\bibfnamefont{V.~A.} \bibnamefont{Soltamov}},
  \bibinfo{author}{\bibfnamefont{E.~N.} \bibnamefont{Mokhov}},
  \bibinfo{author}{\bibfnamefont{P.~G.} \bibnamefont{Baranov}},
  \bibinfo{author}{\bibfnamefont{V.}~\bibnamefont{Dyakonov}},
  \bibinfo{author}{\bibfnamefont{A.}~\bibnamefont{Krueger}},
  \bibnamefont{et~al.}, \bibinfo{journal}{Applied Physics Letters}
  \textbf{\bibinfo{volume}{105}}, \bibinfo{pages}{243112}
  (\bibinfo{year}{2014}).

\end{thebibliography}
\end{document}